\documentclass
[a4,english,twocolumn,aps,nofootinbib,superscriptaddress]{revtex4}%
\usepackage{graphicx}
\usepackage{color}
\usepackage{bm}
\usepackage{amsmath}
\usepackage{amssymb}
\usepackage{amsfonts} 

\begin{document}
\title{Comment on \textquotedblleft Does the weak trace show the past of a quantum
particle?\textquotedblright}
\author{Q.\ Duprey}
\affiliation{ENSEA, 6 avenue du Ponceau, 95014 Cergy-Pontoise cedex, France}
\author{A. Matzkin}
\affiliation{Laboratoire de Physique Th\'eorique et Mod\'elisation, CNRS
Unit\'e 8089, CY Cergy Paris Universit\'e, 95302 Cergy-Pontoise
cedex, France}

\begin{abstract}
In the paper \textquotedblleft Does the weak trace show the past of a quantum
particle?\textquotedblright\lbrack arXiv:2109.14060v2], it is argued that null
weak values of the spatial projectors are inadequate to infer the presence of
a quantum particle at an intermediate time between preparation and detection.
This conclusion relies on two arguments -- (i) the role of the disturbance
induced by a weak measurement, and (ii) classical-like features like
continuous paths that must purportedly be associated with a quantum particle
presence. Here we first show that (i) arises from a misunderstanding of null
weak values by putting forward a simple counter-example that highlights that
the relevant quantities to examine  are the vanishing amplitudes, not the
wavefunction. Then we briefly argue that enforcing classical pre-conditions in
order to account for quantum properties during unitary evolution is unlikely
to lead to a consistent understanding of quantum phenomena.
\newline\newline
\end{abstract}
\maketitle



In order to learn something about a physical system's properties, a
measurement of the system is necessary. For a quantum system, a standard
measurement radically changes the system's evolution as the premeasurement
state is projected to one of the eigenstates of the measured observable. It is
therefore difficult, even in principle, to imagine a procedure that would
enable one to measure the properties of a system at an intermediate time,
without affecting the system's evolution.

With weak measurements \cite{origine} it is possible to achieve minimally
perturbing non-destructive measurements. An obervable $\hat{O}$ of a system
initially prepared (\textquotedblleft preselected\textquotedblright) in state
$\left\vert \psi_{i}\right\rangle $ is weakly measured by coupling the system
to a quantum pointer. \textquotedblleft Weakly\textquotedblright\ means here
that the system's evolution after the coupling is only affected to first
order, so that when a different observable, say $\hat{N}$, is subsequently
measured (through a projective measurement), the probability to obtain the
final state $\left\vert \psi_{f}\right\rangle $ (an eigenstate of $\hat{N}$)
is not modified to first order (relative to the same evolution without the
coupling). Once $\left\vert \psi_{f}\right\rangle $ is obtained
(\textquotedblleft post-selected\textquotedblright), the quantum pointer
coupled to $\hat{O}$ is shifted by $\operatorname{Re}\left(  O^{w}\right)  $
where the weak value $O^{w}$ is given by
\begin{equation}
O^{w}=\frac{\left\langle \psi_{f}\right\vert \hat{O}\left\vert \psi
_{i}\right\rangle }{\left\langle \psi_{f}\right\vert \left.  \psi
_{i}\right\rangle }. \label{w}%
\end{equation}

\begin{figure}[tb]
\centering \includegraphics[width=5cm]{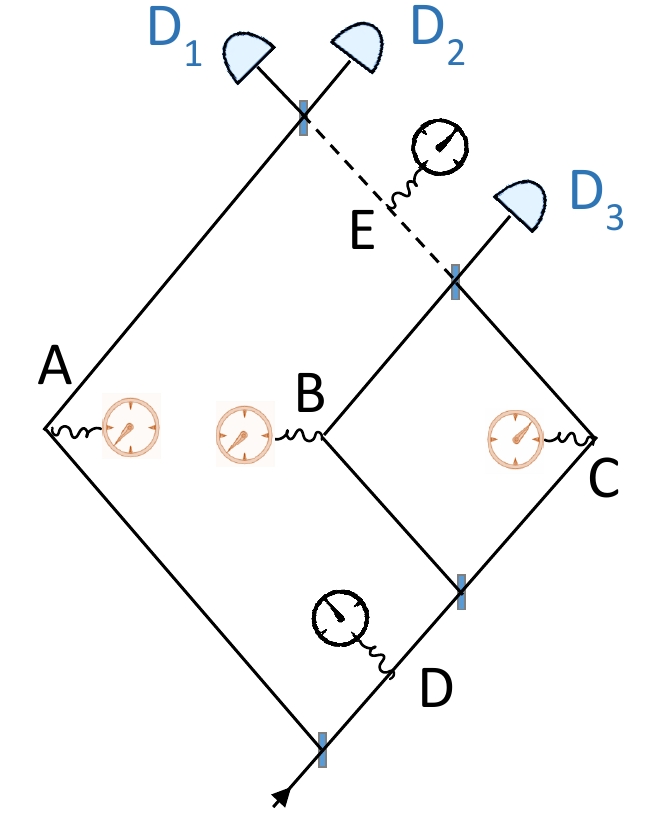}  \caption{The original 3-paths
interferometer with a nested Mach-Zehnder, introduced in \cite{vaidman2013}
and discussed in \cite{P}. We have pictured here the weakly coupled pointers
(the pointers in orange shift after post-selection, while those pictured in
black remain unshifted as the weak value vanishes). The interferometer is
balanced so that destructive
interference is obtained along arm E in the absence of weak interactions.}%
\label{fig1}%
\end{figure}

Let us now examine a quantum particle propagating inside the interferometer
depicted in Fig.\ 1 in a pre and post-selected situation. Weak couplings can
be implemented jointly on the different arms.\ Typically, the corresponding
weak values are generally non-zero and all the coupled quantum pointers will
shift. Vaidman \cite{vaidman2013} proposed that the weak values of the spatial
projector, $\hat{O}=\Pi_{x}\equiv\left\vert x\right\rangle \left\langle
x\right\vert $ could be used as a \textquotedblleft weak trace
criterion\textquotedblright, telling us where the quantum particle has been
during its evolution inside the interferometer. In the typical case just
mentioned, this criterion implies that the particle has been in all the arms
in the interferometer, i.e. a statement of the paths superposition. In some
situations, the weak value for a coupling at position $x_{0}$ will vanish,
$\Pi_{x_{0}}^{w}=0$, indicating that the quantum particle was not there (in
the sense that its spatial degree of freedom was not detected at $x_{0}$).

\begin{figure}[tb]
\centering \includegraphics[width=5cm]{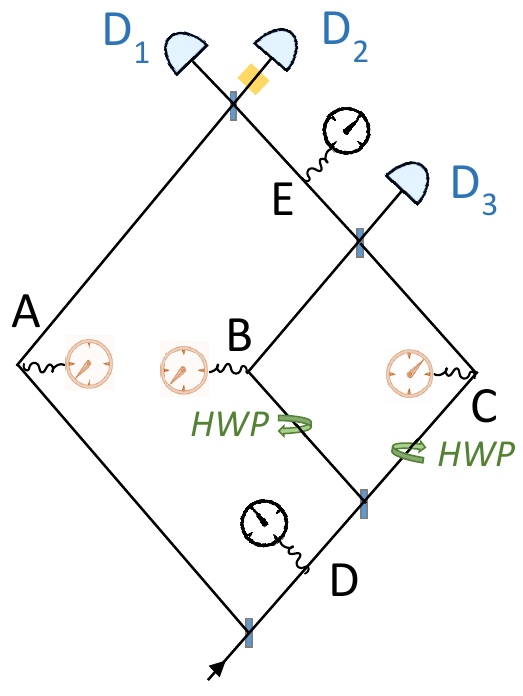}  \caption{Similar to Fig. 1, with
the addition of polarization rotations inside the nested Mach-Zehnder and a
polarizer at post-selection. Now the wavefunction on arm E does not vanish,
although the weak values are the same as those of Fig. 1.}%
\label{fig2}%
\end{figure}

An apparent paradox pointed out by Vaidman \cite{vaidman2013,refs} happens when the
interferometer is balanced such that the wavefunction along arm E
vanishes.\ Then, taking the initial state  $\left\vert \psi_{i}\right\rangle
=\left(  \left\vert \psi_{D}\right\rangle +i\left\vert \psi_{A}\right\rangle
\right)  /\sqrt{2},$ if post-selection is chosen when detector $D_{2}$ clicks,
that is $\left\vert \psi_{f}\right\rangle =\left(  \left\vert \psi
_{A}\right\rangle +i\left\vert \psi_{E}\right\rangle \right)  /\sqrt{2}$, the
following weak values are obtained:%
\begin{align}
\text{ }\Pi_{A}^{w} &  =1\qquad\Pi_{D}^{w}=0\label{c1}\\
\Pi_{B}^{w} &  =\frac{1}{2}\qquad\Pi_{C}^{w}=-\frac{1}{2}\\
\Pi_{E}^{w} &  =0.\label{c3}%
\end{align}
This means that the quantum particle is detected inside the nested
interferometer, but is not detected in the ingoing (D) or outgoing (E) arms.
The weak trace is therefore discontinuous.

In the paper \textquotedblleft Does the weak trace show the past of a quantum
particle?\textquotedblright\ \cite{P} Hance \emph{et al.} claim that the weak
trace approach is inconsistent because weak measurements disturb the system.
They argue that if weak interactions are made inside the nested
interferometer, then the perfect destructive interference on arm E that exists
when no weak couplings are implemented is disturbed, so that the wavefunction
$\psi_{E}(x)$ is not zero anymore. This is indeed the case, but this
observation is irrelevant to having a vanishing weak value.\ Put
differently Eq. (\ref{c3}) may hold irrespective of whether $\psi_{E}(x)$
vanishes or not\footnote{Note this is the case here for arm D, since $\Pi_{D}^{w}=0$ but $\psi_{D}(x)$ is not zero. 
	The post-selected state evolved backward in time does vanish on arm D, but this
	backward evolved state does not seem to be taken as physically real in Ref. \cite{P}.}. A vanishing weak value requires the numerator of Eq.
(\ref{w}), a transition amplitude, to vanish, not the wavefunctions. 

This can be seen by slightly modifying the interferometer pictured in Fig.\ 1
in the following way (see Fig.\ 2). We now include the polarization of the
photon and choose $\left\vert \psi_{i}\right\rangle =\left(  \left\vert
\psi_{D}\right\rangle +i\left\vert \psi_{A}\right\rangle \right)  \left\vert
H\right\rangle /\sqrt{2}$ as the initial state (where $\left\vert
H\right\rangle ,\left\vert V\right\rangle $ stand for horizontal and vertical
polarization, and $\left\vert \nearrow\right\rangle =\left(  \left\vert
H\right\rangle +\left\vert V\right\rangle \right)  /\sqrt{2}$, $\left\vert
\searrow\right\rangle =\left(  \left\vert H\right\rangle -\left\vert
V\right\rangle \right)  /\sqrt{2}$ label the polarization in the diagonal
basis). Inside the nested interferometer, we add wave-plates in order to
rotate the polarization on each arm such that the state of the photon inside
the interferometer becomes $i\left\vert \psi_{B}\right\rangle \left\vert
\nearrow\right\rangle +\left\vert \psi_{C}\right\rangle \left\vert
\searrow\right\rangle $. Now on arm E the wavefunction becomes $\left\vert
\psi_{E}\right\rangle \left(  -\left\vert \nearrow\right\rangle +\left\vert
\searrow\right\rangle \right)  $ which does not vanish. However, it is an easy
exercise to check that the structure of Eqs. (\ref{c1})-(\ref{c3}) is left
untouched -- Eqs. (\ref{c1}) and (\ref{c3}) remain identical while $\Pi
_{B}^{w}$ and $\Pi_{C}^{w}$ pick up a factor depending on the polarization
rotation. $\Pi_{E}^{w}$ vanishes because the state on arm E\ is orthogonal to
the post-selected one.

This counter-example, an adaptation to the present discussion of a 3-paths
atomic interferometer introduced previously \cite{duprey} (see also \cite{soko}), disproves the
argument given in \cite{P} since we still have $\Pi_{E}^{w}=0,$ but now
implementing the weak interactions does not change the state on arm E
from an undisturbed vacuum to a \textquotedblleft perturbed state with light
present\textquotedblright\ \cite{P} (as for any weak measurement, we have a
slight perturbation introduced by the weak couplings). And the trace of the
spatial degree of freedom of the quantum particle remains discontinuous. Note
that in this counter-example the post-selected state evolved backward in time does not vanish in arm D, although 
$\Pi_{D}^{w}=0$.

The second aim of \cite{P} is to show that \textquotedblleft the weak trace
does not reveal the path of a quantum particle\textquotedblright. The authors
assert that a quantum particle must have a continuous path (like a classical
particle), as per their condition ii). As it is given, this condition appears
somewhat arbitrary and not particularly meaningful: it depends on how a path
is defined for a quantum particle, and has no relation with an observational
warrant that could confirm this claim. Indeed, the wavefunction is continuous
and can be understood as propagating along Feynman paths.\ But these paths can
interfere, and the destructive interference between amplitudes carried by
different Feynman paths is precisely what makes a weak value vanish
\cite{APRR}.\ However the wavefunction or the Feynman paths are usually taken
to be mere computational tools (at least according to standard quantum
mechanics), so while it is possible to explain discontinuous weak value traces
in terms of continuous but destructively interfering paths, this hardly
changes the experimental fact that position weak values are discontinuous.

The underlying issue here is not to decree that quantum particles \emph{must}
have continuous paths, but (a) to put forward a cogent framework in order to
define properties of a quantum system at an intermediate time; and (b) suggest
an experimental scheme in order to observe such properties.\ Obviously, one can refuse
to ascribe properties to quantum systems at intermediate times, but then the path of 
a quantum particle cannot even be defined (and the issue of whether these paths are continuous or not becomes moot).
It turns out that
weak measurements already constitute a framework in which weak values can be
related to quantum properties, provided one is willing to relax the
eigenstate-eigenvalue link (see \cite{FP} and Refs.\ therein). And the
resulting weak values, even if they predict a non-classical behavior, can be
experimentally observed.

To conclude, we have shown that the arguments given in \cite{P} aiming to show
that weak values of the spatial projector give an inconsistent account of the
past position of a quantum particle are incorrect. The interpretation of weak
values is controversial \cite{controversy}, and it is interesting to confront
different viewpoints in order to understand the elusive nature and properties
of quantum systems. Nevertheless, it is preferable to base the discussion on
technically relevant arguments.

\end{document}